# Advantages of the coherent state compared with squeeze state in unidimensional continuous variable quantum key distribution

Xuyang Wang[1,2,*], Yanxia Cao[1], Pu Wang[1], and Yongmin Li[1,2,*]

[1]*State Key Laboratory of Quantum Optics and Quantum Optics Devices, Institute of Opto-Electronics, Shanxi University, Taiyuan 030006, China*
[2]*Collaborative Innovation Center of Extreme Optics, Shanxi University, Taiyuan 030006, China*
*\*wangxuyang@sxu.edu.cn*
*\*yongmin@sxu.edu.cn*

In this work, a comparison study between unidimensional (UD) coherent-state and UD squeeze-state protocols is performed in the continuous variable quantum key distribution domain. First, the UD squeeze-state protocol is proposed, and the equivalence between the prepare-and-measure and entanglement-based schemes of UD squeeze-state protocol is proved. Then, the security of the UD squeeze-state protocol under collective attack in realistic conditions is analyzed. Lastly, the performances of the two UD protocols are analyzed. Based on the uniform expressions established in our study, the squeeze-state and coherent-state protocols can be analyzed simultaneously. Our results show that the UD squeeze-state protocols are quite different from the two-dimensional (TD) protocols in that the UD squeeze-state protocols have a poorer performance compared with UD coherent-state protocols, which is opposite in the case of TD protocols.

## I. INTRODUCTION

The unconditional security of quantum key distribution (QKD) prevents information from being eavesdropped; it is expected that this technology will be used for a wide variety of applications in the future with the advent of quantum information technology. In general, QKD technology can be categorized into discrete-variable and continuous-variable (CV) QKD protocols [1–2]. CV-QKD protocols encode information into continuous quadrature components of quantum states and utilize homodyne detectors instead of single-photon detectors. The CV-QKD protocols typically provide a high secret key rate at a relatively short distance; in addition, they have good compatibility with the classical communication networks [3-25].

Based on the utilized quantum states, the CV-QKD protocols can be classified as coherent-state protocols, squeeze-state protocols, and entanglement-state protocols. In general, it is believed that the squeeze-state and entanglement-state protocols perform better than the coherent-state protocols; however, because coherent state sources are easy to prepare, coherent-state protocols have also been widely researched. Till date, many protocols to enhance the performance of or simplify a QKD system have been proposed, including a unidimensional (UD) coherent-state protocol [26], three coherent-state protocols [27], and method for passive state preparation [28]. The advantages of the UD coherent-state protocol include easy modulation, low costs, and only needs less random numbers [29, 30]. In the case of a small amount of excess noise, the UD coherent-state protocol performs as well as the two-dimensional (TD) coherent-state protocol (GG02). Therefore, the UD coherent-state protocol has the potential to be used for applications in various scenarios, such as in QKD local area networks, where the transmission distance between users is typically short and cost is a key concern.

Thus far, the UD modulation method has only focused on the coherent-state protocol. Therefore, in this study, a UD squeeze-state protocol was proposed; in addition, the equivalence between the prepare-and-measure (PM) scheme and entanglement-based (EB) scheme of the protocol was proved. Using the uniform expression introduced in our study for the squeezed and coherent states, we can analyze the protocols of the states simultaneously. In particular, by changing the value of a parameter $r$, the result for the corresponding state can be obtained. By comparing these obtained results, we observed that in the case of the UD protocols, the coherent-state protocol performed better that the squeeze-state protocol, whereas, in the case of the TD or symmetrical modulation protocols, the asymmetric squeezed state led to a better performance.

The remainder of this paper is organized as follows. In Section II, the proposed UD squeeze-state protocol is presented; in addition, the equivalence of PM and EB schemes is discussed. Section III presents the security analysis of the UD squeezed-state protocol under collective attack in realistic conditions using the EB scheme. Further, Section

IV discusses the performance of the UD squeeze-state protocol and compares the protocol with the UD coherent-state protocol. Finally, our conclusions are provided in Section V.

## II. UD SQUEEZED AND COHERENT PROTOCOLS

### A. PM scheme for UD squeezed or coherent state protocols

It is well-known that one of the quadrature variances of the squeezed state is less than the shot noise, whereas the other quadrature variance is more than the shot noise. When the amplitude quadrature, which is denoted by $x$, is squeezed, the covariance matrix is given by:

$$\gamma_x = \begin{pmatrix} e^{-2s} & 0 \\ 0 & e^{2s} \end{pmatrix}, \tag{1}$$

where $s > 0$ is the squeezing parameter. When the phase quadrature, which is denoted by $y$, is squeezed, the covariance matrix is given by:

$$\gamma_y = \begin{pmatrix} e^{2s} & 0 \\ 0 & e^{-2s} \end{pmatrix}. \tag{2}$$

The covariance matrix of the coherent state is

$$\gamma_c = \begin{pmatrix} 1 & 0 \\ 0 & 1 \end{pmatrix}. \tag{3}$$

In order to describe the three kinds of states uniformly in our study, the covariance matrix was denoted as follows:

$$\gamma = \begin{pmatrix} 1/r & 0 \\ 0 & r \end{pmatrix}, \tag{4}$$

where $r$ represents the variance of the phase quadrature. When $0 < r < 1$, the matrix represents the covariance matrix of the phase quadrature squeezed state or $y$-squeezed state, whereas when $r > 1$, the matrix represents the covariance matrix of the amplitude quadrature squeezed state or $x$-squeezed state. Further, when $r = 1$, the matrix is the covariance matrix of the coherent state. It should be noted that all the variances in our study are normalized to the shot noise. Based on this uniform expression, we can analyze these protocols simultaneously.

The traditional TD squeeze-state protocol in the PM scheme proposed in [5] is based on the following concept: Alice randomly prepares $x$-squeezed states displaced along $x$ or $y$-squeezed states displaced along $y$ with a Gaussian distribution. Then, the mixed Gaussian states with the null mean value and covariance matrix are obtained as follows:

$$\gamma_{sym} = \begin{pmatrix} V_M + e^{-2s} & 0 \\ 0 & e^{2s} \end{pmatrix} = \begin{pmatrix} e^{2s} & 0 \\ 0 & V_M + e^{-2s} \end{pmatrix} = \begin{pmatrix} V & 0 \\ 0 & V \end{pmatrix}, \tag{5}$$

where $V_M$ is the modulation variance. When we impose $e^{-2s} + V_M = e^{2s} = V$, a thermal state with variance $V$ can be obtained. The thermal state realized by a mixture of $x$-squeezed states is indistinguishable from a thermal state realized by a mixture of $y$-squeezed stated. Thus, the information can be encoded in two conjugate quadratures with both mixed states representing the same thermal state of variance $V$.

Similar to the UD coherent-state protocol, for the UD squeezed-state protocol in the PM scheme, Alice displaces the squeezed state along one quadrature with a Gaussian distribution. Without loss of generality, the amplitude quadrature is selected to distribute the squeezed state as shown in Fig. 1. The squeezed state can be either the $y$-squeezed state as shown in Fig. 1 (a) or $x$- squeezed state as shown in Fig. 1 (b). Figure 2 (c) presents the scheme of the UD coherent-state protocol. The modulation variance along the amplitude quadrature is $V_M$. Further, Bob measures the amplitude or phase quadratures by switching the detection bases randomly with true random numbers.

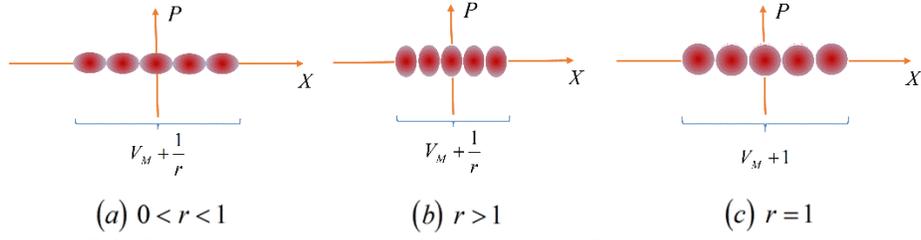

FIG. 1 UD distributed squeezed states in the phase space. (a) Phase quadrature squeezed state. (b) Amplitude quadrature squeezed state. (c) Coherent state.

After Bob has measured all the pulses, the two partners need to perform post-processing, which starts by applying sifting. In our study, we select the sifting method corresponding to reverse reconciliation. The sifting steps are described as follows:

1. Bob discloses the random measurement base of each pulse.
2. Alice records the data corresponding to the cases wherein Bob measures the amplitude quadrature. It should be noted that Bob should store all the data. The amplitude quadrature data are used to estimate the parameters and distill the secret key. Furthermore, the phase quadrature data are used to estimate the variance of phase quadrature in order to calculate the secret key.

Then. they make public part of the amplitude quadrature randomly to estimate the parameters, such as transmission efficiency and excess noise in amplitude quadrature, after which, the secret key rate is calculated. The procedure of reverse reconciliation and privacy amplification is used to ensure that Alice and Bob could share a group of secret keys.

### B. Equivalence of the PM and EB schemes in UD protocols

In general, most of the current experimental systems for CV-QKD protocols are based on PM schemes, because they are easy to implement in practice. However, in theory, it is difficult to analyze the security of such protocols based on the PM schemes. On the contrary, theoretical analysis based on the EB scheme can be performed appropriately; the involved entangled states lead to simpler and more feasible calculations [2]. In particular, in the case of the UD protocol, security analysis based on the EB scheme has more advantages that are based on the PM scheme. The covariance matrices obtained using the EB schemes contain the constraints of phase amplitude quadrature; however, these constraints are difficult to obtain using the PM scheme.

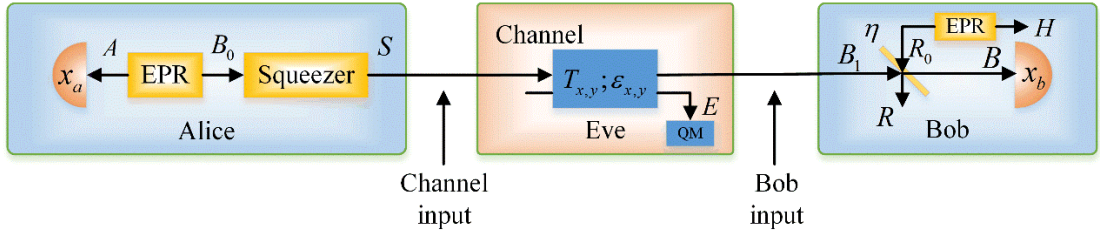

FIG. 2. Schematic of the EB scheme of the UD protocol under realistic conditions.

In the EB scheme shown in Fig. 2, Alice prepares a pair of Einstein–Podolsky–Rosen (EPR) states $\rho_{AB_0}$ with covariance matrix $\gamma_{AB_0}$ as follows:

$$\gamma_{AB_0} = \begin{pmatrix} V \cdot I_2 & \sqrt{V^2-1} \cdot \sigma_z \\ \sqrt{V^2-1} \cdot \sigma_z & V \cdot I_2 \end{pmatrix}, \tag{6}$$

where $I_2 = diag(1,1)$ and $\sigma_z = diag(1,-1)$. Then, Alice squeezes one of its modes $B_0$ with the squeeze parameter $S = \ln\sqrt{V/r}$. The resulting covariance matrix $\gamma_{AS}$ is:

$$\gamma_{AS} = (I \oplus SQ)\gamma_{AB_0}(I \oplus SQ)^T$$

$$= \begin{bmatrix} V & 0 & \sqrt{V(V^2-1)/r} & 0 \\ 0 & V & 0 & -\sqrt{r(V^2-1)/V} \\ \sqrt{V(V^2-1)/r} & 0 & V^2/r & 0 \\ 0 & -\sqrt{r(V^2-1)/V} & 0 & r \end{bmatrix} = \begin{bmatrix} \gamma_A & \sigma_{AS} \\ \sigma_{AS}^T & \gamma_S \end{bmatrix}, \quad (7)$$

where $SQ$ is the squeeze operator, which is given by:

$$SQ = \begin{bmatrix} \sqrt{V/r} & 0 \\ 0 & 1/\sqrt{V/r} \end{bmatrix}. \quad (8)$$

Here, $\gamma_A$ and $\gamma_S$ are the covariance matrices of the modes $A$ and $S$, respectively, and $\sigma_{AS}$ is the correlation matrix of the two modes. In quantum communication, Alice sends the mode $S$ to Bob. The collapsed state $\rho_s$ that transmitted to Bob depends on the measurement of mode $A$. In the UD protocol, when the modulation is performed on the amplitude quadrature in the PM scheme, Alice will conduct homodyne detection on the amplitude quadrature in the EB scheme. The covariance matrix of mode S conditioned on Alice's measurement result $x_A$ can be derived using:

$$\gamma_S^{x_A} = \gamma_S - \sigma_{AS}(X \cdot \gamma_A \cdot X)^{MP}\sigma_{AS}^T, \quad (9)$$

where $X = diag(1,0)$, and MP denotes the Moore–Penrose inverse of a matrix. After a straightforward calculation, we can obtain the following:

$$\gamma_S^{x_A} = \begin{pmatrix} 1/r & 0 \\ 0 & r \end{pmatrix}. \quad (10)$$

Before Alice's measurement, the two modes of state $\rho_{AS}$ are centered on $d_A^{in} = (0,0)$ and $d_S^{in} = (0,0)$. The homodyne measurement on mode $A$, denoted by $m = (x_A, 0)$, will project the mode $S$ to the squeezed state or coherent state centered on:

$$d_S^{out} = \sigma_{AS}(X \cdot \gamma_A \cdot X)^{MP}(m - d_A^{in}) + d_S^{in}. \quad (11)$$

Then, after a straightforward calculation, we can obtain:

$$d_S^{out} = \sqrt{(V^2-1)/rV} \cdot (x_A, 0). \quad (12)$$

From the covariance matrix $\gamma_A$ of mode $A$, we can observe that the variance of $x_A$ is $V$. Therefore, for the amplitude quadrature, the variance of the center value of mode $S$ is:

$$Var(d_S^{out}) = \frac{V^2-1}{rV} \cdot Var(x_A) = \frac{V^2}{r} - \frac{1}{r}. \quad (13)$$

Because the variance of each collapsed state in the amplitude quadrature is $1/r$ (the first diagonal element of Eq. (10)); therefore, the total variance of the squeezed state in the amplitude quadrature is $V^2/r$ (which is obtained by adding the result of Eq. (13) and $1/r$). This corresponds to the modulation variance $V_M = (V^2-1)/r$ in the PM scheme; in addition, the variance of the modulated state is $1/r$. Thus, we can see that the two schemes are indistinguishable for Bob and Eve, i.e., they are equivalent. The advantageous consequence of this equivalence is that the experiment can be performed using the PM scheme, whereas its security can be studied using the EB scheme.

### III. SECURITY ANALYSIS OF THE UD COHERENT AND SQUEEZED STATE PROTOCOLS UNDER REALISTIC CONDITIONS

In the previous section, we established the equivalence between the EB and PM schemes of the UD protocols. Here, we analyze the security of the protocols to primarily study the availability of the UD squeeze-state protocol. The secret key rate against collective attacks for reverse reconciliation in the asymptotic regime can be calculated as follows:

$$\Delta I = \beta \cdot I_{AB} - \chi_{BE}, \qquad (14)$$

where $\beta$ is the reverse reconciliation efficiency; thus far, the highest value achieved is 99.96% [31]. $I_{AB}$ is the Shannon mutual information between Alice and Bob, while $\chi_{BE}$ is the Holevo bound, which represents the maximum information eavesdropped by Eve under collective attacks.

$I_{AB}$ can be calculated using Shannon's equation as follows:

$$I_{AB} = \frac{1}{2} \log_2 \frac{V_A}{V_{A|B}}, \qquad (15)$$

where $V_A$ is the variance of Alice in the case of amplitude quadrature; its value can be found at the first diagonal element of covariance matrix $\gamma_{AB_1}$, which is given as follows:

$$\gamma_{AB_1} = \begin{bmatrix} \sqrt{1+rV_M} & 0 & \sqrt{T_x V_M}(1+rV_M)^{1/4} & 0 \\ 0 & \sqrt{1+rV_M} & 0 & C_y^{B_1} \\ \sqrt{T_x V_M}(1+rV_M)^{1/4} & 0 & T_x(V_M + 1/r + \chi_{\text{linex}}) & 0 \\ 0 & C_y^{B_1} & 0 & V_y^{B_1} \end{bmatrix}, \qquad (16)$$

where $\chi_{\text{linex}} = (1-T_x)/T_x + \varepsilon_x$ is the channel noise added relative to the channel input in amplitude quadrature, $(1-T_x)/T_x$ is the noise due to losses, and $\varepsilon_x$ is the excess noise in the amplitude quadrature. $C_y^{B_1}$ is the unknown correlation of phase quadratures, and $V_y^{B_1}$, which could be measured experimentally, is the variance of the phase quadrature. The conditional variance $V_{A|B}$ is the first diagonal element of the conditional matrix $\gamma_{A|B}$, which can be derived as follows:

$$\gamma_{A|B} = \gamma_A - \sigma_{AB}(X\gamma_B X)^{MP}\sigma_{AB}^T, \qquad (17)$$

where $X = \text{diag}(1,0)$, $\gamma_A$, $\gamma_B$, and $\sigma_{AB}$ are all submatrices of the covariance matrix $\gamma_{AB}$. All of these are shown below.

$$\gamma_{AB} = \begin{bmatrix} \sqrt{1+rV_M} & 0 & \sqrt{\eta T_x V_M}(1+rV_M)^{1/4} & 0 \\ 0 & \sqrt{1+rV_M} & 0 & C_y^{B_1}\sqrt{\eta} \\ \sqrt{\eta T_x V_M}(1+rV_M)^{1/4} & 0 & \eta T_x(V_M + 1/r + \chi_{\text{totx}}) & 0 \\ 0 & C_y^{B_1}\sqrt{\eta} & 0 & \eta(V_y^{B_1} + \chi_{\text{hom}}) \end{bmatrix} = \begin{bmatrix} \gamma_A & \sigma_{AB} \\ \sigma_{AB}^T & \gamma_B \end{bmatrix} \qquad (18)$$

with $\chi_{hom} = (1+v_{el})/\eta - 1$ is the noise introduced by the realistic homodyne detector relative to Bob's input in the amplitude quadrature, and $\chi_{totx} = \chi_{linex} + \chi_{hom}/T_x$ is the total noise added between Alice and Bob relative to the channel input in the amplitude quadrature. $v_{el}$ is the electronic noise of the homodyne detector. Finally, the Shannon mutual information can be derived as follows:

$$I_{AB} = \frac{1}{2} \log_2 \left( \frac{1/r + V_M + \chi_{tot}}{1/r + \chi_{tot}} \right). \tag{19}$$

To obtain the covariance matrix $\gamma_{AB_1}$ which is the submatrix of $\gamma_{AB_1 FE}$, it is convenient to assume that Eve holds a purification of Alice's and Bob's mutual state $\rho_{AB_1}$. Considering the freedom-in-purification theorem, any purification of $\rho_{AB_1}$ that Eve may possess will result in the same entropy and hence the same Holevo information $\chi_{BE}$ [32]. Here, we suppose that Eve generates an EPR state $\rho_{EE'}$ with covariance matrix $\gamma_{EE'}$ and replaces the channel with a lossless channel in which she inserts a beam splitter with phase sensitive transmission $T_x$ and $T_y$ (as shown in Fig. 3). The beam splitter mixes the modes $S$ and $E'$. Then, Eve retains one of the output mode $F$ for herself and passes the other mode $B_1$ to Bob. This process can be expressed as follows:

$$\gamma_{AB_1 FE} = (I_2 \oplus BS_{SE'} \oplus I_2) \cdot (\gamma_{AS} \oplus \gamma_{E'E}) \cdot (I_2 \oplus BS_{SE'} \oplus I_2)^T. \tag{20}$$

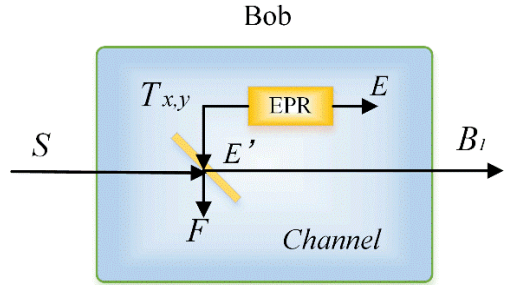

FIG. 3. Beam splitter mode of the channel.

The covariance matrix $\gamma_{ABRH}$ can be obtained using a similar procedure shown below:

$$\gamma_{ABRH} = (I_2 \oplus BS_{B_1 R_0} \oplus I_2) \cdot (\gamma_{AB_1} \oplus \gamma_{R_0 H}) \cdot (I_2 \oplus BS_{B_1 R_0} \oplus I_2)^T. \tag{21}$$

In order to analyze the security of the protocol in an easy manner, the realistic homodyne detector is usually replaced by a beam splitter $BS_{B_1 R_0}$ with a transmission efficiency $\eta$ and ideal homodyne detector (as shown in Fig. 2). The electronic noise $v_{el}$ of the homodyne detector can be modeled by one model $R_0$ of ERP state $\rho_{R_0 H}$ with variance $V_N = 1 + v_{el}/(1-\eta)$ entering the other input port of the beam splitter. Because the detector cannot be accessed by the eavesdropper, it is considered that the detector has phase insensitive efficiency.

The Holevo bound that represents the maximum information eavesdropped by Eve is defined as:

$$\chi_{BE} = S(\rho_{FE}) - S\left(\rho_{FE}^{x_b}\right). \tag{22}$$

Because the quantum state $\rho_{AB_1 FE}$ and $\rho_{ARHFE}^{x_b}$ are all pure states, $S(\rho_{AB_1}) = S(\rho_{FE})$ and $S(\rho_{EF}^{x_b}) = S(\rho_{ARH}^{x_b})$. We can rewrite $\chi_{BE}$ as:

$$\chi_{BE} = S(\rho_{AB_1}) - S(\rho_{ARH}^{x_b}), \tag{23}$$

where $S(\rho)$ is the von Neumann entropy of the quantum state $\rho$. For an $n$-mode Gaussian state $\rho$, this entropy can be calculated using the symplectic eigenvalues of the covariance matrix $\gamma$ characterizing $\rho$ as follows:

$$S(\rho) = \sum_i G\left(\frac{\lambda_i - 1}{2}\right), \tag{24}$$

where $G(x) = (x+1)\log_2(x+1) - x\log_2 x$. In general, the symplectic eigenvalues of covariance matrix $\gamma$ with $n$ mode can be calculated by finding the absolute eigenvalues of the matrix $i\Omega\gamma$. The matrix $\Omega$ has an expression as

$$\Omega = \bigoplus_{k=1}^{n} \begin{bmatrix} 0 & 1 \\ -1 & 0 \end{bmatrix}. \tag{25}$$

The covariance matrix $\gamma_{ARH}^{x_b}$ of the state $\rho_{ARH}^{x_b}$ can be derived using Eq. (26).

$$\gamma_{ARH}^{x_b} = \gamma_{ARH} - \sigma_{ARH;B}(X\gamma_B X)^{MP}\sigma_{ARH;B}^T, \tag{26}$$

where $\gamma_{ARH}$, $\gamma_B$, and $\sigma_{ARH;B}^T$ are all submatrices of the matrix $\gamma_{ARHB}$ as shown below:

$$\gamma_{ARHB} = \begin{bmatrix} \gamma_{ARH} & \sigma_{ARH;B} \\ \sigma_{ARH;B}^T & \gamma_B \end{bmatrix}. \tag{27}$$

The covariance matrix $\gamma_{ARHB}$ is obtained by rearranging the lines and columns of matrix $\gamma_{ABRH}$, which, in turn, is obtained using Eq. (21). In the above expression (23), there are two unknown variables $C_y^{B_1}$ and $V_y^{B_1}$. In order to obtain the secret key rate, we need to constraint them using the covariance matrix $\gamma_{AB_1}$ and Heisenberg uncertainty principle as follows [33]:

$$\gamma_{AB_1} + i\Omega \geq 0. \tag{28}$$

Then, the following parabolic equation can be derived:

$$(C_y^{B_1} - C_0)^2 \leq \frac{V^2 - 1}{V} \frac{\chi_{\text{linex}}}{1/r + \chi_{\text{linex}}}(V_y^{B_1} - V_0), \tag{29}$$

where

$$C_0 = -\frac{\sqrt{(V^2-1)/r}}{\sqrt{T_x V}(1/r + \chi_{\text{line}})} \quad \text{and} \quad V_0 = \frac{1}{T_x(1/r + \chi_{\text{linex}})}. \tag{30}$$

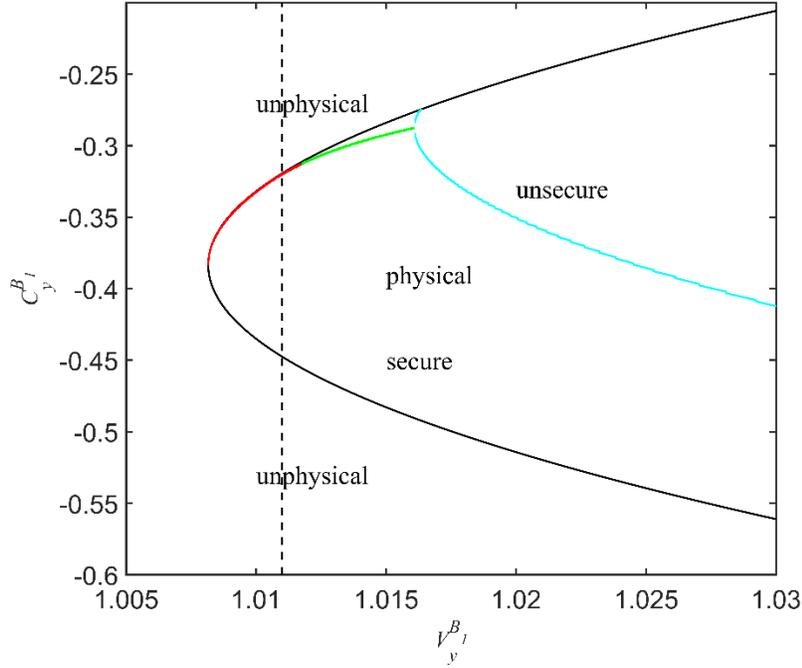

FIG. 4. Regions of the UD 1-dB $x$-squeezed state protocol under realistic detection conditions.

In Fig. 4, a black parabolic curve between $C_y^{B_1}$ and $V_y^{B_1}$ can be seen. The parameter $r$ is set to 1.1, which indicates that 1-dB $x$-squeezed states are generated. The values of $r$ correspond to the coherent state or $y$-squeezed state, which has a similar result. The other parameters are set as follows: $\beta = 0.99$, $V_M = 3$, $T_x = 0.1$, $\varepsilon_x = 0.01$, $\eta = 0.6$, and $\upsilon_{el} = 0.1$. The entire plane is divided into physical and unphysical regions by the parabolic curve. In particular, the physical region is contained in the parabolic curve. In the unphysical region, the value of $C_y^{B_1}$ and $V_y^{B_1}$ cannot be satisfied simultaneously. The cyan curve with the secret key rate of zero separates the entire physical region into secure and unsecure regions. The secret key rate in the secure region is larger than zero, whereas it is less than zero in the unsecured region. For a fixed value $V_y^{B_1}$, there is a group of secret key rates with different values of $C_y^{B_1}$. In general, the value of $C_{y\min}^{B_1}$, which can be achieved by scanning all the $C_y^{B_1}$ in the secure region, corresponding to the minimum secret key rate $\Delta I_{\min}$ is used for securing. When the value of $V_y^{B_1}$ is different, the value of $C_{y\min}^{B_1}$ will be different. The line with two colors, i.e., red and green, record the trajectory of $C_{y\min}^{B_1}$ in the secure region. In contrast, the red part indicates the points with the minimum secret key rate that overlap with points on the parabolic curve. When the value of $V_y^{B_1}$ increases, the blue part represents that the points with the minimum secret key rate gradually separate from the parabolic curve. We denote the line as "safe line." The minimum secret key rate of safe line is shown in Fig. 5.

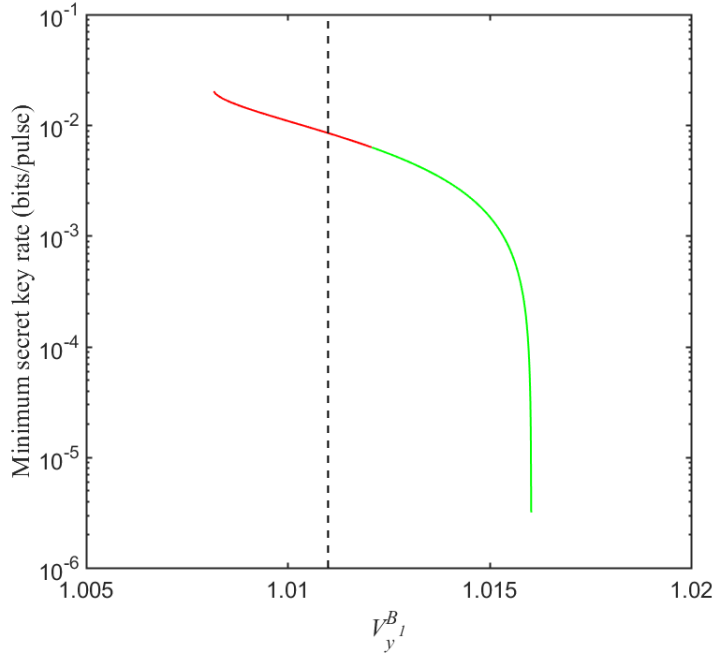

FIG. 5. The minimum secret key rate of safe line.

When the transmission efficiency $T_y$ in the phase quadrature equals the transmission quadrature $T_x$ in the amplitude quadrature, the variance of phase quadrature $V_y^{B_1}$ can be derived as follows:

$$V_y^{B_1}\big|_{T_x=T_y} = T_x\left(r + \chi_{line}\right) \tag{31}$$

It is depicted as the vertical black virtual line in Figs. 4 and 5. The minimum secret key rate at the phase quadrature variance $V_y^{B_1}\big|_{T_x=T_y}$ is typically used to estimate the secret key rate of the realistic conditions. Thus, we denote the virtual line as the "expected line".

Based on the above discussion, we can observe that the security of the squeeze- and coherent-state protocols can be analyzed simultaneously using uniform expressions.

## IV.   PERFORMANCE OF THE COHERENT AND SQUEEZE STATE PROTOCOLS

In order to analyze the performance of the coherent- and squeeze-state protocols, the secret key rate $\Delta I_{min}$ versus distance for different squeezing parameters were plotted, as shown in Fig. 5. The black solid line corresponds to the squeezed parameter $r = 1$, indicating the coherent state. The squeezed parameters of the blue dash and blue dot lines are $r = 0.8$ and $r = 0.6$, respectively, both of which represent the amplitude quadrature squeezed state. The squeezed parameters of the red dash-dot and red dash-dot-dot lines are $r = 1.2$ and $r = 1.4$, respectively, both of which represent the phase quadrature squeezed state. The other parameters are the same as the ones used in Fig. 4.

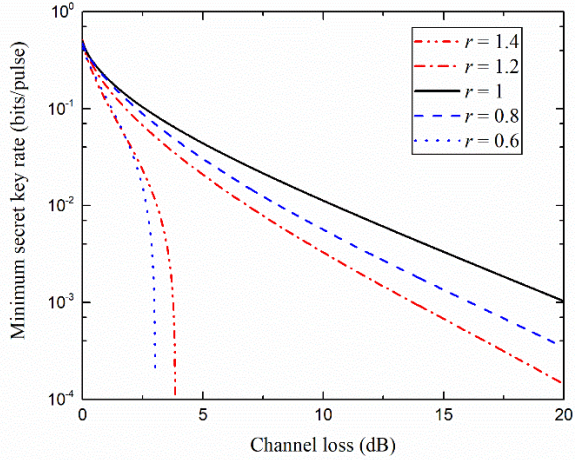

FIG. 6. Minimum secret key rate versus distance at different squeezing parameters.

Based on an analysis of the performance of the UD protocols, we can observe that either the UD amplitude quadrature or phase quadrature squeezed state has a lower performance. In general, the larger the degree of squeezing, the lower the performance is. This observation is quite different from the TD or symmetrical protocols in which the squeeze-state protocol performs better than the coherent-state protocol; in that case, the performance increases with an increase in the degree of squeezed parameters [2, 34].

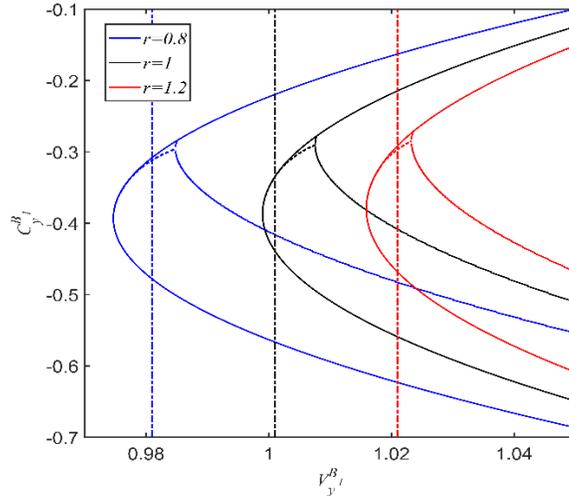

FIG. 7. Parabolic curves with different squeezing parameters.

The parabolic curves with different squeezing parameters are shown in Fig. 7. With an increase in the value of the squeezed parameters, the parabolic curve moves from left to right. The cross point of the safe line (dash line) and expected line (dash-dot line) also moves from left to right. The minimum secret key rates for different conditions are graphically depicted in Fig. 8. Evidently, the black cross point that represents the coherent state protocol has the highest secret key rate. In particular, many of the black cross points at different transmission efficiencies constitute the black solid line that is shown in Fig. 6; similarly, the blue cross and red cross points at different transmission efficiencies constitute the corresponding blue dash and red dash-dot lines in Fig. 6.

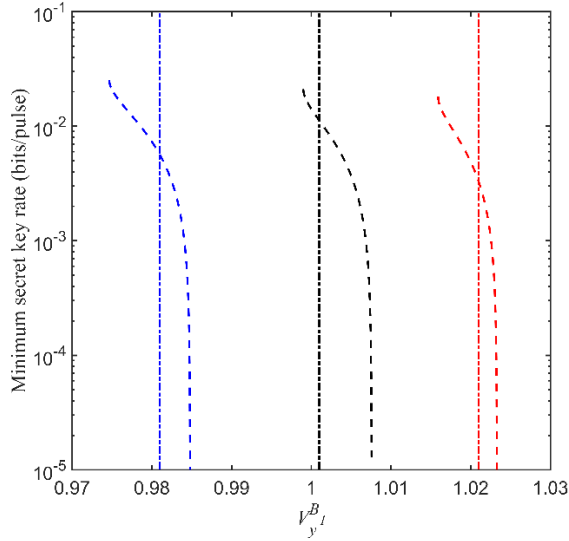
FIG. 8. Secret key rates of different squeezing parameters.

## V.     DISCUSSION AND CONCLUSIONS

By comparing the performance of the UD protocols, we can see that the performance of the UD squeeze-state is lower than that of the UD coherent-state protocols. It is different from the TD protocols. In this section, we will discuss the reasons briefly and present a conclusion.

It is well known that the Shannon mutual information of the TD squeeze-state protocol between Alice and Bob is

$$I_{AB}^{TDS} = \frac{1}{2}\log_2\left(\frac{V+\chi_{tot}}{1/V+\chi_{tot}}\right) \tag{32}$$

and the Shannon mutual information of the coherent-state protocol between the two parties is

$$I_{AB}^{TDC} = \frac{1}{2}\log_2\left(\frac{V+\chi_{tot}}{1+\chi_{tot}}\right) \tag{33}$$

where $V$ is the variance of EPR state in the equivalent EB scheme, and it is larger than one [35]. When the value of $V$ is same, we can see that the TD squeeze-state protocol has a larger Shannon mutual information than the TD coherent state protocol by comparing expressions (32) and (33). In the equivalent EB scheme, the information eavesdropped by Eve in the two TD protocols can be calculated by

$$\chi_{BE} = S(\rho_E) - S(\rho_E^{x_b}). \tag{34}$$

It means that when the variance of the EPR state is the same, Eve eavesdrops the same information in the above two TD protocols. Using expression (14), we can obtain

$$\Delta I_{AB}^{TDS} > \Delta I_{AB}^{TDC}.$$

In UD protocols, the Shannon mutual information $I_{AB}^{UD}$ between Alice and Bob can be calculated by expression (19). It can be rewritten as

$$I_{AB}^{UD} = \frac{1}{2}\log_2\left(\frac{V^2/r+\chi_{tot}}{1/r+\chi_{tot}}\right). \tag{35}$$

The value of $I_{AB}^{UD}$ not only changes with variance $V$ but also changes with variable $r$. The variable $r$ has a relation with the squeezing process, and it can be used to discriminate the UD protocols. However, $I_{AB}^{UD}$ is monotonous with the value of $r$. If the value of $r$ is larger, then the value of $I_{AB}^{UD}$ is larger. Thus, we cannot determine the superiority of the UD coherent-state protocol by only using the Shannon mutual information briefly. Because of the uncertain variables $C_y^{B_1}$, the Holevo bound $\chi_{BE}^{UD}$ can be calculated directly. The Minimum secret key rate should be achieved firstly by scanning the variable $C_y^{B_1}$. When the value $C_{y\min}^{B_1}$ is determined, the Holevo bound $\chi_{AB}^{UD}$ can be calculated. Thus, it is hard to use the monotonicity of $I_{AB}^{UD}$ and $\chi_{BE}^{UD}$ to determine the superiority of UD coherent-state protocol.

From the final numerically calculated result of minimum secret key rate shown in Fig. 6 and Fig. 8, we can see that when $r=1$ or when the coherent state is used, the best performance can be achieved. This is in contrast to the trend in the case of the TD protocols. Because of our analysis, the coherent state is proven to be the optimal state in the UD domain. For future research and experiment, an integration of CV QKD in deployed optical-network-based UD coherent-state protocol is expected [36], especially, when the transmission distance between users is usually short and cost is a key concern. In theory, the composable security [37, 38] of the UD coherent-state protocol will be considered.

## ACKNOWLEDGMENTS

This research was supported by the Key Project of the Ministry of Science and Technology of China (2016YFA0301403), National Natural Science Foundation of China (NSFC) (11504219, 61378010), Shanxi 1331KSC, and Program for the Outstanding Innovative Teams of Higher Learning Institutions of Shanxi.